\documentstyle{article}

\font\elevenbf=cmbx10 scaled\magstep 1
\font\elevenrm=cmr10 scaled\magstep 1
\font\elevenit=cmti10 scaled\magstep 1

\newcommand{\beq}{\begin{equation}}
\newcommand{\beqn}{\begin{eqnarray}}
\newcommand{\eeq}{\end{equation}}
\newcommand{\eeqn}{\end{eqnarray}}
\def\slash#1{\setbox0=\hbox{$#1$}#1\hskip-\wd0\hbox to\wd0{\hss\sl/\/\hss}}
\renewenvironment{thebibliography}[1]
 { \elevenrm
   \begin{list}{\arabic{enumi}.}
    {\usecounter{enumi} \setlength{\parsep}{0pt}
     \setlength{\itemsep}{3pt} \settowidth{\labelwidth}{#1.}
     \sloppy
    }}{\end{list}}

\begin{document}
\begin{center}{{\elevenbf INFRARED CORRECTIONS TO THE TOP QUARK WIDTH\\}
\vglue 1.0cm
{\elevenrm Pankaj Jain, Douglas W. McKay and Herman J. Munczek \\}
\vglue 0.7cm
\baselineskip=13pt
{\elevenit  Department of Physics and Astronomy \\}
\baselineskip=12pt
{\elevenit The University of Kansas\\}
\baselineskip=12pt
{\elevenit Lawrence, KS 66045-2151\\}
\vglue 2.0cm
{\elevenrm ABSTRACT}}
\end{center}
\vglue 0.3cm
{\rightskip=3pc
 \leftskip=3pc
 \elevenrm\baselineskip=12pt
 \noindent
 We present a nonperturbative analysis of the
top quark propagator
using the Schwinger Dyson equation in the ladder approximation
including both the electroweak as well as the infrared QCD effects.
 We find that
the infrared effects are
negligible only for top mass larger than about 250 GeV.}
\vfill
\noindent
Presented at the DPF92 meeting, Fermilab, Chicago, Nov. 10-14, 1992
\eject
{\elevenbf\noindent 1. Introduction}
\vglue 0.1cm
\baselineskip=14pt
\elevenrm
The success of perturbative QCD is based on the fact that the
QCD coupling gets smaller at large momentum. Therefore for any process
for which the dominant contribution comes from the high momentum
region, perturbative QCD should yield quantitatively reliable predictions.
However for several interesting cases,
perturbation theory fails to give reasonable results. For example, for the
calculation of bound state masses the perturbative gluon propagator
does not give correct results even
for quark masses much larger then the typical scale of nonperturbative
QCD. Presumably the infrared region gives a significant contribution
in this case.
Another important example is quark confinement. The fact
that quarks are never observed as free particles suggests that
the quark propagator is considerably different from the perturbative
propagator and might receive significant contributions from the
infrared region. This is supported by nonperturbative calculations, performed
by choosing phenomenologically and theoretically motivated
models for gluon propagator, which suggest that, independently of the
mass of the quark, the dressed quark
propagator is very different from the free propagator and in
particular admits no mass pole, presumably a signal of confinement.
Intuitively we can argue that, because of the large coupling, a
quark has a large amplitude to exchange a soft gluon with itself
which can in principle modify its singularity structure.
This argument may not hold for the top quark because of
its large decay rate into a W boson and a bottom quark. This
suggests the possibility that the top quark might decay before
exchanging a soft gluon with itself. If this were the case then
the infrared region will be completely ignorable and perturbation
theory becomes quantitatively reliable.

To get a rough idea of the mass of the top at which the infrared
effects become negligible we compare the width of the top with the
typical hadronic scale. If the width is much smaller than the
scale of confinement then the top will have sufficient time to
exchange a soft gluon. For a top quark of mass 150 GeV, its purely
electroweak width at tree level is about 800 MeV. This is clearly
comparable to the hadronic scale and we cannot rule out the possility
of significant infrared corrections. For the top quark of mass 250 GeV,
however, the top width is larger than 4 GeV and the infrared effects
should be completely negligible.

The top propagator can be calculated nonperturbatively by solving the
Schwinger-Dyson (SD) equation [1],
\beq
S^{-1}(q) = S^{-1}_0(q) - i \int {d^4k\over (2\pi)^4} \gamma_\mu S(k)
\gamma_\nu G_{\mu\nu}(k-q)
\eeq
We employ the ladder approximation for our calculation and choose
phenomenologically and theoretically motivated models for the gluon
propagator.
We choose the Landau gauge for our calculation, in which case
the gluon propagator has the form,
$G_{\mu\nu}(k) = (g_{\mu\nu}-k_\mu k_\nu/k^2)G(k^2)$.
We are interested in calculating the corrections to the top
width due to the low momentum region and therefore need a
model for $G(k^2)$ which might be qualitatively reliable
in this region. There is some theoretical and phenomenological
evidence that the gluon propagator has a $1/k^4$ behavior
at small momentum. We therefore choose the following two models
for $G(k^2)$, which represent a regularized form of the
$1/k^4$ behavior.
\beq
G^a(k) = i{4\over 3}(2\pi)^4\eta^2\delta^4(k)
\eeq
\beq
G^b(k) = {8\pi a\over (k^2-\epsilon^2)^2} .
\eeq
 The second model $G^b(k)$ leads to a linear potential, $V(r)=ar$, in
the nonrelativistic limit as long as $\epsilon$ is small compared to the
scale of confinement. We calculate the parameters in these potentials by
fitting the light and heavy meson spectrum and decay constants [1],
which give a range of values for these parameters.
\vglue 0.6cm
{\elevenbf\noindent 2. Results and discussion}
\vglue 0.1cm
\baselineskip=14pt
\elevenrm
In this section we present the results for the top width obtained by
choosing the two gluon propagator model given in Eqns. 2 and 3.
Since we are only interested in getting a qualitative idea, the
parameters in these potentials will be chosen within their range
such that they give the
smallest correction to the tree level calculation of width.
For the potential $G^a$, this implies that $\eta=450$ MeV and for
$G^b$, we get a=(450 MeV)$^2$, $\epsilon=150$ MeV.
The numerical results are given in Table 1.

\vglue 0.3cm
{\rightskip=3pc
\leftskip=3pc
\noindent
Table 1: Results for the top quark width $\Gamma$ including infrared QCD
corrections for the gluon propagator models given in Eqns. 2 and 3.
$\Gamma_0$ is the top width at tree level.
\vglue 0.3cm}
\begin{tabular}{|c|c|c|c|}
\hline
$m_t$ (GeV) & $\Gamma_0$ (GeV) & $\Gamma$ (GeV) & $\Gamma$ (GeV)\\
 & &(model $G^a$) & (model $G^b$)\\
\hline
100 & 0.0922 & 0.0461 & 0.0715\\
150 & 0.80 & 0.40 & 0.61\\
190 & 2.0 & 1.44 & 1.85 \\
234 & 4.0 & 3.79 & 3.91  \\
\hline
\end{tabular}
\vglue 0.3cm
Our results show that the low momentum region gives a negligible
correction to the top width only if its mass is larger than
about 250 GeV. For a top quark of mass 150 GeV, the infrared
region can give a significant correction. These results
follow the trend, as anticipated in the introduction,
 that as the width of the top gets much
larger than the scale of confinement the infrared region
becomes negligible. For comparison, perturbative QCD calculations
gives about a 10\% correction [2].

In conclusion, we have shown here that infrared effects can give
 significant contribution to the top width if the mass of the
top quark is less than about 150 GeV. However once the mass gets
larger than about 250 GeV, the infrared corrections are
completely negligible. The magnitude of the corrections is
dependent on the width of the top. The corrections get
negligible only if the width is much larger than the
scale of confinement.

\vglue 0.5cm
{\elevenbf \noindent 3. Acknowledgements \hfil}
\vglue 0.4cm
We acknowledge useful discussions with John Ralston. This work was
supported in part by the Department of Energy under grant No.
DE-FG02-85-ER40214.
\vglue 0.5cm
{\elevenbf\noindent 4. References \hfil}
\vglue 0.4cm

\end{document}